\documentclass[manuscript]{aastex}

\accepted{to the PASP}

\def\mbar{\ifmmode\overline{m}\else$\overline{m}$\fi}
\def\Mbar{\ifmmode\overline{M}\else$\overline{M}$\fi}
\def\mibar{\ifmmode\overline{m}_I\else$\overline{m}_I$\fi}
\def\MIbar{\ifmmode\overline{M}_I\else$\overline{M}_I$\fi}
\def\Nbar{\ifmmode\overline{N}\else$\overline{N}$\fi}

\def\psf{\ifmmode_{\hbox{PSF}}\else PSF\fi}
\def\deg{\ifmmode^\circ\else$^\circ$\fi}
\def\arcsec{\ifmmode^{\prime\prime}\else$^{\prime\prime}$\fi}
\def\arcmin{\ifmmode^{\prime}\else$^{\prime}$\fi}

\def\dmod{\ifmmode(m{-}M)_0\else$(m{-}M)_0$\fi}
\def\mM{\ifmmode(m{-}M)_0\else$(m{-}M)_0$\fi}
\def\vi{\ifmmode(V{-}I)\else$(V{-}I)$\fi}
\def\viz{\ifmmode(V{-}I)_0\else$(V{-}I)_0$\fi}
\def\EBV{\ifmmode E_{B-V}\else$E_{B-V}$\fi}

\shorttitle{Variability in NGC 2301}
\shortauthors{Tonry et al.}
\received{16 Nov. 2004}

\begin{document}
\title{A Search for Variable Stars and Planetary Occultations in NGC2301 I:
Techniques}

\author{John L. Tonry,\altaffilmark{1}
Steve Howell,\altaffilmark{2}
Mark E. Everett,\altaffilmark{3}
Steven A. Rodney,\altaffilmark{1}
Mark Willman\altaffilmark{1}
and Cassandra VanOutryve\altaffilmark{2,4}
}

\altaffiltext{1}{Institute for Astronomy, University of Hawaii,
2680 Woodlawn Drive, Honolulu, HI 96822; {jt@ifa.hawaii.edu},
{rodney@ifa.hawaii.edu}} 
\altaffiltext{2}{WIYN Observatory and NOAO, 950 N. Cherry Ave.,
Tucson, AZ  85726; {howell@noao.edu}}
\altaffiltext{3}{Planetary Science Institute, Fort Lowell Rd., Tucson, AZ
85719; {everett@psi.edu}}
\altaffiltext{4}{Current address: Astronomy Department, University of 
California, Berkeley, CA 94720; {cassy@berkeley.edu}}

\begin{abstract}

We observed the young open cluster NGC 2301 for 14 nights in Feb. 2004
using the orthogonal transfer CCD camera (OPTIC).
We used PSF shaping techniques (``square stars")
during the observations allowing
a larger dynamic range (4.5 magnitudes) of high photometric precision
results ($\le$2 mmag) to be obtained. These results are better than
similar observing campaigns using standard CCD imagers.
This paper discusses our observational techniques and presents initial results
for the variability statistics found in NGC 2301. Details of the variability
statistics as functions of color, variability type, stellar type, 
and cluster location will appear in paper II.
\end{abstract}


\newpage

\section{Introduction}
The study of stellar variability has long been a part of astronomical
research and is the mainstay for understanding stellar evolution,
stellar formation, stellar death, and galactic dynamics. Variable stars
are fundamental to such astrophysical pursuits as composition of the ISM,
evolution of galaxy clusters, and cosmological distance scales. Two areas of
variability have been ignored by most astronomers using large telescopes
(due to lack of sufficient observing
time, lack of the proper instrumentation,  and/or 
insufficient photometric precision); low amplitude variations
($\sim$0.1 mag or less) and frequent time sampling for extended periods
(i.e., few minute sampling for many days in a row). Studies of these two  
regions of phase space, when tied together with large areal surveys or
dense stellar regions, are also capable of providing detections and limits
on the number and type of transiting extra-solar planets encircling
solar-type stars.

An initial study reaching high photometric precision and large area
was presented in Everett et al., (2002) and Everett \& Howell (2001). The
latter paper also presents a brief review of other similar variability studies.
The Everett et al., study
concentrated on field stars covering a large range of color and luminosity
class. Here, we present a similar high photometric precision study of
0.15 sq. degrees centered on the young, metal rich star cluster NCG 2301.
This paper outlines the novel observational technique\footnote{The technique
used here is described in detail in Howell et al., 2003.} 
we used and details the
new data reduction and analysis applied to the nearly 5000 (mostly
solar-type stars)
light curves we collected over a 14 night period in Feb. 2004. Paper II
will provide details of the stars themselves, their light curves, and
a discussion of how this study reveals new insights into open clusters 
and NGC 2301 itself.

\section{Observations}
\subsection{High Precision Photometry}
The accuracy of any observation is ultimately limited by 
photon statistics (assuming one can control other factors, especially
systematic noise), which
for stellar sources is approximately the peak count in electrons per
pixel times the area of the point spread function (\psf).  Typical
CCDs can collect 30,000 to 100,000~e$^-$ per pixel 
prior to the onset of non-linearity
or full well, and our desired accuracy of 1~millimag (mmag) per exposure can
therefore be achieved with a \psf\ which is at least 3--6~pixels FWHM.
However, if we are trying to simultaneously observe many stars over a wide
range of apparent magnitude
with a precision of 1~mmag, we must collect
$10^6$~e$^-$ from the faintest stars without saturating the brightest
ones.  For example, if we want to maintain precision over 5
magnitudes, we require a \psf\ of FWHM 30-60~pixels.  Of course this
incurs a penalty of increased photon statistics from sky noise, and
increased systematic error from overlapping \psf\ but these
considerations are factored into the decision about the optimal \psf\
size to use.

One way of achieving the above scenario 
is to defocus the telescope, but this causes
difficulties because the ``donut'' \psf\ is not constant across the
image, causing problems with photometry from \psf\ fitting which is
essential because of overlapping stellar images.  Additionally,
defocused images often place power in very extended, asymmetric wings
of the \psf\ and highlight optical imperfections.  Creating a
large \psf\ in order to sample as much of the stellar
apparent magnitude range as possible, sampling as large a field of view as
possible, and at the same time as trying to achieve extreme (mmag)
photometric accuracy are conflicting requirements with defocussed
images.  Obviously the quantitative level one can attain with
defocussed images depends on the specifics of the telescope, detector,
cleanliness of optics, etc, but at the UH2.2-m telescope, for example, there is
enough vignetting over our 10\arcmin\ field of view that trying
to fit defocussed images with a constant \psf\ leads to systematic
errors well in excess of 1~mmag.  We have chosen to exploit the
capabilities of orthogonal transfer CCDs and rapidly raster scan the
stellar images over a square \psf\ during the course of an exposure.
Initial work on this type of ``square star" photometry is discussed in
Howell et al., (2003).  We were not completely successful in reaching
photon-limited performance for reasons discussed below, but we believe
that this technique could do so with some refinement.

Square star production allows us to control the size and shape of the extended
\psf\ however, they are not a perfect solution to all types of observing. For
example, field crowding can become an issue, very faint background
stars incur additional uncertainty from increased sky signal, and
two-dimensional objects, such as galaxies, are rendered unusable for
most purposes.  Also, the square star resides in more pixels than a
traditional round star would, thereby bringing additional read noise
(and dark current if an issue) into play. This increased noise is not
of real concern here however, as we are working in the regime of
stellar photon noise being the dominant source of uncertainty.

\subsection{The Field}

There is an optimal stellar density to survey many stars at once.
Obviously a very low stellar density nets us very few stars per
exposure, whereas a very high density causes us problems from
overlapping images.  A simple calculation reveals that the greatest
number of non-overlapping stellar images which are found at random
locations is $N/(en)$, where $N$ is the number of pixels in the
detector, $n$ is the number of pixels in the \psf, and $e$ is the base
of natural logarithms.  Of course the real situation is more
complicated because we can tolerate overlaps to some degree and a
range of stars of very different brightnesses
which affect the ability to extract photometry from close stars. Even
so, this indicates that for a camera such as OPTIC, which has $4096^2$
pixels and a \psf\ which we fix at $30^2$ pixels, to sample 5
magnitudes of stellar brightness, we ought to aim for a stellar
density which gives us at least 10,000 stars over the full range of apparent
magnitude of interest.
This is a higher density than we could actually achieve,
given the rather small pixel size ($0.138$\arcsec). (We had hoped to
employ a focal reducer at the 2.2-m telescope, but it was not ready in
time.)  Obviously we could push the density of stars with $10^6$
photons as high as we like by extending the observing time, but this
has to be traded off against the sampling frequency of several times
per hour which we wanted to search for planets.

We searched the USNO-B catalog over those right ascensions available
for our time allocation of first half nights in February, looking for
fields which offered high stellar density, good visibility over most
of the night, and slanted toward a high likelihood field of view to
detect transiting planets. We choose to center our survey on the young
(164 million years), open Galactic cluster NGC~2301.  This cluster is
of high metallicity (+0.06), 872 pc away, and has a low reddening
(0.028 mag). The distance modulus to NGC~2301 is 9.79 making the
turnoff magnitude (near main sequence spectral type A0) occur at R=10,
or just below our saturation limit. We therefore did not collect valid
photometric time series data on any evolved higher mass main sequence
stars (i.e., giants) but did find two stars that have apparently
already passed into the white dwarf stage.  It would have been
possible to reach a higher density in the center of a globular
cluster, but not over multiple 10\arcmin\ fields of view, and we did
not want to look for planets in such an environment.

\subsection{The Observing Run}

\begin{figure}[t]
\epsscale{1.00}
\plotone{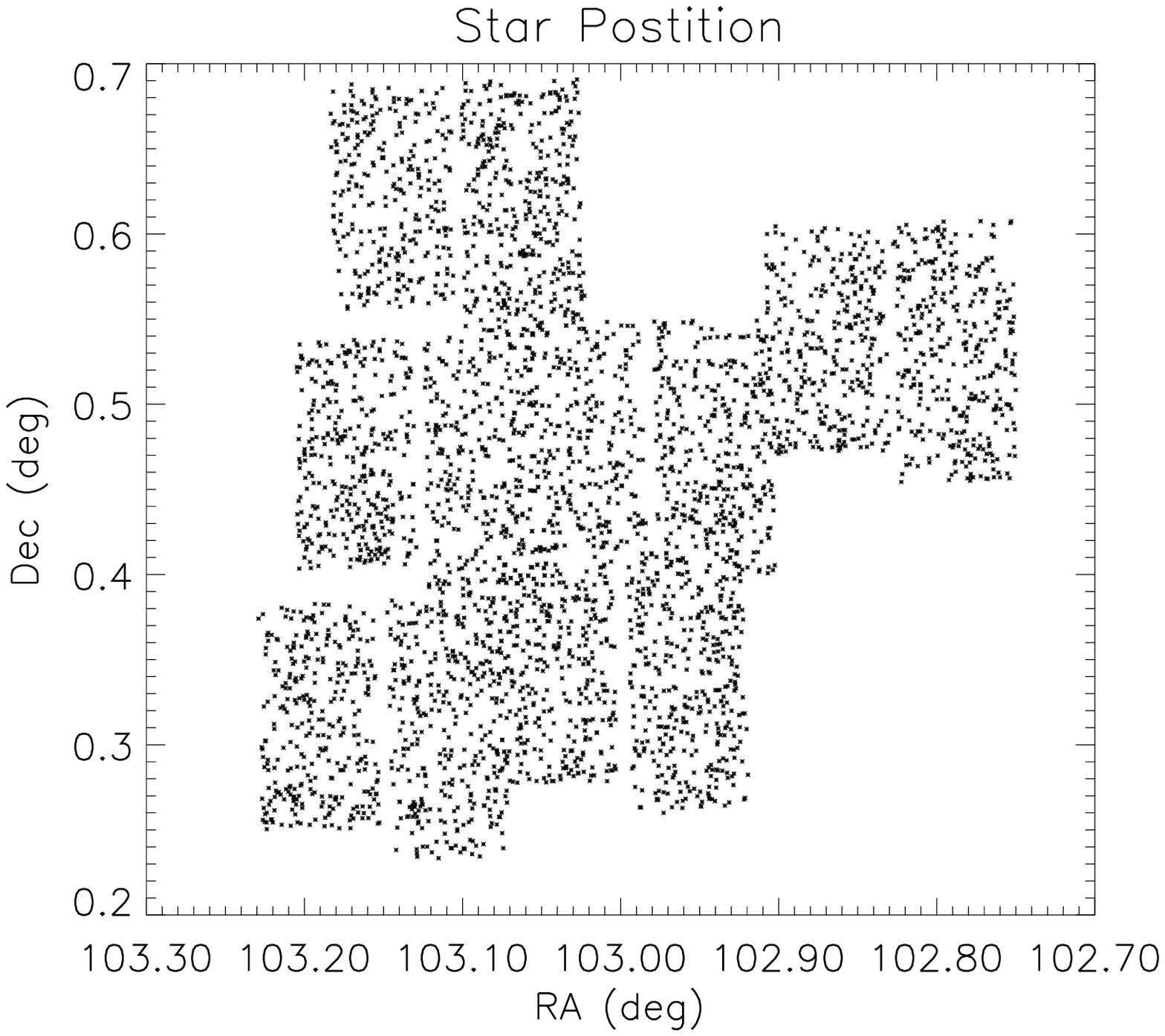}
\caption{The six OPTIC fields in the cluster are shown by 
plotting the locations of the stars in this study. There is some
overlap in the fields and we designate the fields as 
C, N, S, E, W, and SE (for Center, 
North, etc.) We guided each exposure using the lower left on-chip guide region 
in each field which accounts for the slight asymmetry in each OPTIC field. 
\label{sixfields}}
\end{figure}

We were granted the first half of 14 nights at the UH~2.2-m telescope
atop Mauna Kea by the UH TAC: 09-Feb-2004
through 22-Feb-2004.  Two of these nights were completely lost to
bad weather, two were curtailed from the time-series program to collect
calibrated photometry (B,R) and photometric calibration information,
but the remainder of the time
was spent moving rapidly between six adjacent fields around NGC~2301
(see Figure~\ref{sixfields}). A
typical cycle would involve scripted, automatic telescope
offset to the first field, acquisition of guide star,
a 120~s exposure in $R$ band,
approximately 60~s for readout and offset to the next field, for a net
duty cycle of 2/3.  We therefore cycled through the six fields in
approximately 16 minutes, revisiting each star that often.  Since we
were allocated half nights, we could manage 5--6 hours on target,
depending on how quickly after sunset we were willing to start (our
first observations of each night generally show distinctly worse
residuals because of the brighter sky), and we accumulated an average
of 17 cycles over all six fields on each of our 10 full nights.  

The observations were made using the OPTIC camera, the prototype and only
operating orthogonal transfer CCD camera. OTCCDs allow the collected charge
to be moved within the array in response to either on-chip guide stars
or at the user's whim in pre-determined patterns. Based on the experience
gained in our previous observations with these types of CCDs and OPTIC 
(Tonry et al. 1997; Howell et al. 2003), we decided to undertake a
comprehensive variability study using shaped stars. While promising in
initial tests, the challenges of using this new technique in a
relatively crowded field, over many nights and converting our reduction
techniques to deal with square stars while keeping high photometric 
precisions were non-trivial.

The observations were all taken with a $30\times30$ pixel raster
pattern applied to the accumulating charge in the CCD, relative to the
light from the stars which was kept fixed on the CCD by guiding on a
selected star in an unshifted guide region.  This raster scan was
carried out 5 times during the 120~s exposure, with a typical guide
rate of 12~Hz, so the 900 pixels within the $30\times30$ raster
pattern were swept over in about 288 steps, amounting to approximately
a 3~pixel ($\sim$ 0.5 FWHM) charge shift per guide iteration.  See
Howell et al. (2003) for details.

In addition to the square star observations, we also took focused,
non-rastered images in $B$ and $R$ with exposure times of 2, 5, 10,
15, 50, 100, 150, and 300 seconds to establish astrometric and
photometric measures for our stars.  After flattening, these images
were each run through the DoPhot program (Schechter et al. 1993) to
identify the stars present and measure their fluxes.
Another program then took the DoPhot output files and matched
up all the stars between pairs of images (some pairs with extremely
different exposure time had no overlap).  The median magnitude
difference was calculated for each pair, the ensemble of differences
placed in an antisymmetric matrix, the distribution quartiles in an
accompanying error matrix, and the whole solved for a vector of
photometric offsets between the different exposures.  This enabled us
to get relative magnitudes in $B$ and $R$ between $10<R<20.5$ and
$10<B<22)$ with internal error less than 0.01 mag. 
Comparison of the applied
R time series magnitudes to constant stars in the field overlap regions
also showed that we obtained internal photometric calibrations
across fields to 0.01-0.02 magnitudes for our brightest 6 magnitudes of stars. 
On a relatively photometric night we observed Rubin-152 at a variety
of airmasses, which established our photometry zero points in $B$ and
$R$ to about 0.025 mag, and permitted us to put all our B and R
magnitudes were then set on a standard photometric scale.

\begin{figure}[t]
\epsscale{1.00}
\plotone{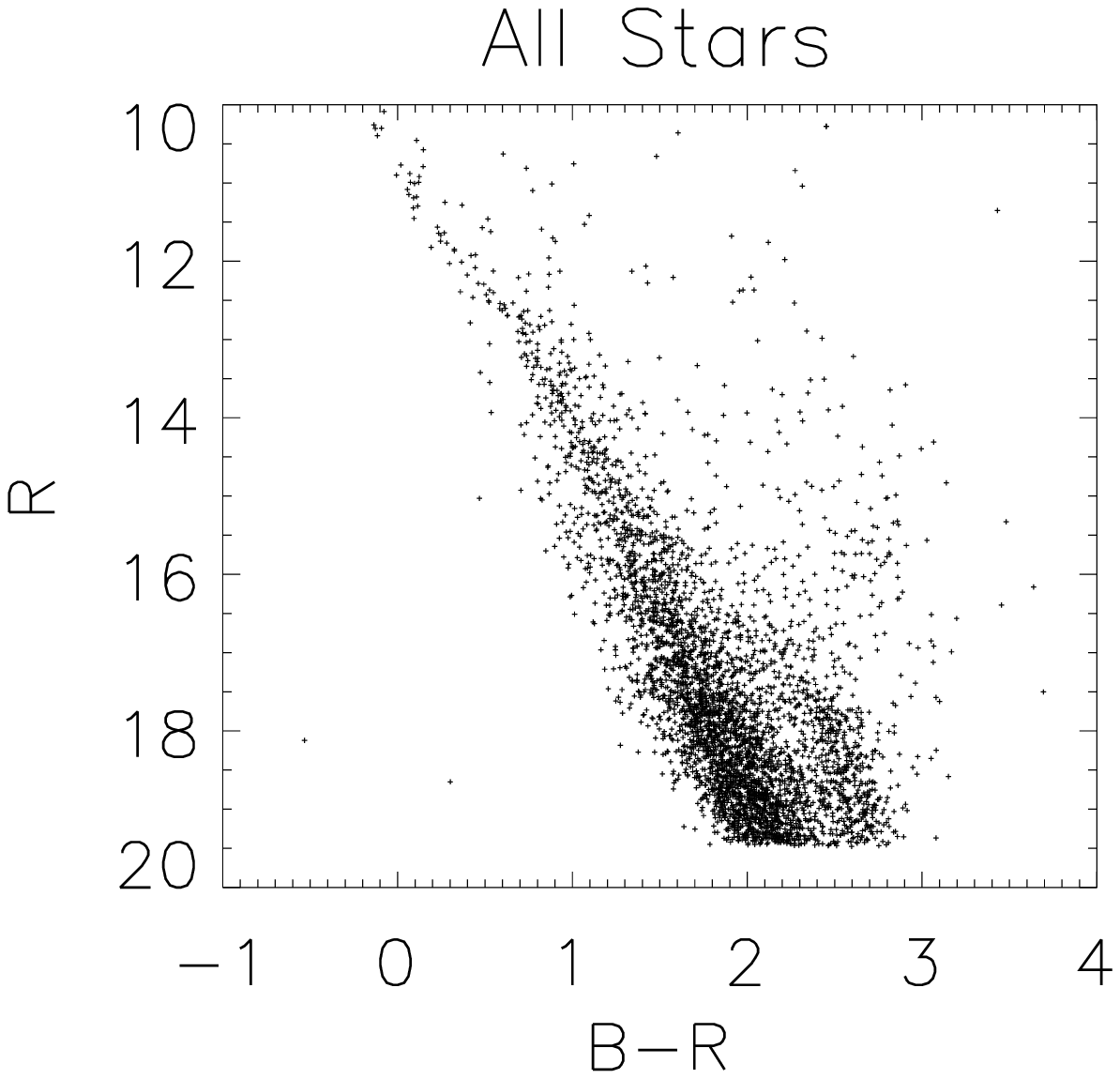}
\caption{Color magnitude diagram for NGC2301. This CMD combines all our stars
from the six fields. Note the broadening of the main sequence due to 
non-cluster member contamination (blue side below R$\sim$16.5) 
and the lower pre-main
sequence stars lying above the ZAMS. The brightest stars we
have valid data for are near the cluster turn-off at an absolute magnitude near
0.
\label{allcmd}}
\end{figure}

Figure~\ref{allcmd} shows our color-magnitude diagram (CMD) for NGC~2301.
The main sequence for NGC~2301 is very apparent and broadened due to binaries
within the main sequence stars. Our  time series
observations, which saturate at $R<10$, do not quite reach the cluster
turnoff at R$\sim$10 but do reveal 
2 new white dwarf candidates. This is a somewhat surprising result
due to the young age of this cluster as only stars earlier than about A0
can have evolved this far. Spectroscopic and cluster membership confirmation
for these sources are planned. 
The lower right hand corner of Figure~\ref{allcmd} shows that a
number of stars (late K and M) are still evolving to the main sequence
in this young cluster. We also note the dense distribution of stars lying below
and to the left of the main sequence starting near $R\sim16.5$ which are most
likely non-cluster contaminates.

In order to provide positions for each of our program objects, we
performed astrometry using our non-rastered images (i.e., normal type CCD
images with round stars taken with the
same camera and telescope and used for the B,R calibration).  Calculating
astrometric positions for the stars was straightforward.  We extracted
stars from the USNO-B catalog within 50\arcmin\ of our field center,
which distilled down to about 1700 stars encompassing each of the six
fields.  OPTIC has two CCDs, so we matched the DoPhot coordinates for
both CCDs independently in each of the six fields to the USNO-B stars
using Brian Schmidt's ``starmatch'' program.  A simple cubic fit to
the stars with matched CCD and USNO-B coordinates then gave us the
means to get from CCD coordinates to RA and Dec.  As usual, the
coordinates we generate therefore have the high relative precision of
CCD astrometry and the good, averaged absolute precision of the USNO-B
catalog.  We expect that the relative positional uncertainty of our
coordinates is something like 0.1 pixel (15 mas) since all of our
stars are bright, and the absolute positional uncertainty is limited
by USNO-B at something like 100 mas.

\subsection{The Reductions}

Data taken with orthogonal shifting are smoother than conventional CCD
data because each bit of sky is integrated on a set of CCD pixels.
However, this means that each observation must have a custom flat field
constructed by convolving an unshifted flat field with the shift
pattern of each image.  
This was done for all observations, and each bias-subtracted
data image was divided by its flat field.

Using the very deep, complete list of stars from the photometry
observations, we assembled an empirical \psf\ for each image by
selecting ``reasonably isolated'' stars and adding their images
together and subtracting the overlaps from nearby stars.
Specifically, any star which has a neighbor brighter than half its
flux closer than 100 pixels, or is contaminated by masked pixels, or
has a mean flux greater than 30,000 or less than 10,000~ADU per pixel,
or is too close to the edge of the chip was rejected.  The rest
contributed to a linear fit between mean brightness of the \psf\
``mesa'' and flux expected from the photometry file.  This fit was
analyzed to identify stars which were saturated or otherwise outliers,
and the remainder form the list of final contributors.  The fit was
also used to set the scale factor between DoPhot fluxes and star
fluxes in this observation.  The square image created by each
contributor star was shifted by the closest integer offset into
alignment and a 
subarray of size $\approx60$ pixels around each added.  The \psf\ from
the previous iteration was then multiplied by the tabulated fluxes for
all stars, and any overlapping data from the contributing stars was
subtracted.  This procedure was then iterated 10 times.

Experiments reveal that the fractional pixel position errors are an
insignificant contributor to error in the eventual fluxes which are
derived from these \psf\ subarrays.  However, we encountered an
unanticipated problem which limits our photometry to about 1~mmag at
best and more like 2~mmag typically.  The problem is that we shift the
charge relative to the stars at a rate of about 5\arcsec/sec, and this
unfortunately means that we are moving the seeing disk of the stellar
images by about the seeing diameter during the time the seeing changes
size and shape and scintillates due to atmospheric turbulence.  Faster
shifting would smear out these effects; slower shifting would cause
them to integrate out more.  As it is, our \psf\ mesas have slightly
corrugated tops and these corrugations change markedly across the
9.2\arcmin\ field of view, although they correlate well for nearby
stars (within $\sim$2 arc min). The average \psf\ we have assembled
does not match the individual stars perfectly, leaving behind
seeing-sized bumps which are at a level of a few $10^{-4}$ of the
original flux.  There are perhaps 30--50 independent 0.6--0.8\arcsec\
bumps across the $30\times30$ pixel ($4\times4$\arcsec) \psf\ which
leads to an error which is of order $10^{-3}$.  In principle it would
be simple to blur out this effect by shifting {\it much} faster during
the exposure (we could potentially work at 2000\arcsec/sec).  We could
also do a more careful data analysis which would involve using local
\psf\ stars (within $\sim$2 arc-minutes) to analyze their own
neighborhoods (see Everett \& Howell 2001).  For the purpose of
finding extra-solar Jupiters causing 15~mmag extinction and a variable
star census, however, we are temporarily satisfied with a 1-2~mmag
error floor.

Once an image has a \psf\ subarray extracted, the flux analysis is
relatively simple.  Stars which are fainter than some limit according
to the photometry list are simply removed from the image by scaling
the \psf\ to their flux and subtracting.  This limit was 300~$e-$/sec
summed over the entire \psf\ for the flux, which is approximately
$R=19.4$, depending on conditions.  There are typically 600--700 stars
remaining in each of our fields brighter than this, and we do a simultaneous
least squares fit for the brightness and sky level of each star.
The fit produces quantitative covariances for overlapping stars, of
course, but to date we have ignored this information except to delete
stars with more than 10\% neighbor contamination from the variability
statistics below.  We would be happy if uncertainty were limited by
the photon statistics, but this does not happen for stars which are
brighter than about $R\sim17$ or 3000~e$^-$/sec total (i.e. for which the
uncertainties should be less than 1.7~mmag).  We therefore estimate an
uncertainty for each star by examining the rms residual after
subtracting the \psf\ fit, and multiplying by the square root of the
number of independent seeing footprints within the square \psf.

\begin{figure}[t]
\epsscale{1.00}
\plotone{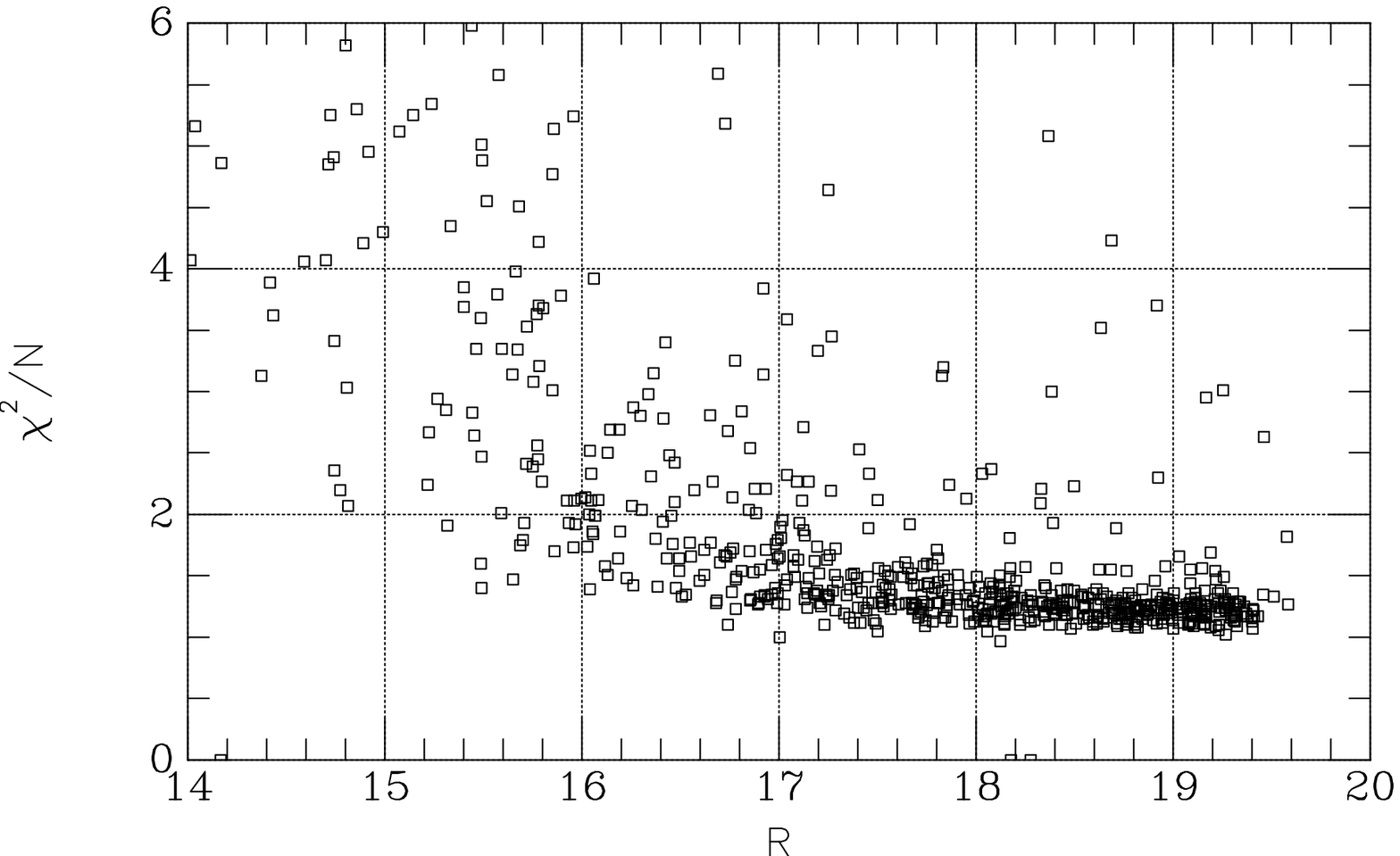}
\caption{$\chi^2/N$ for stars from a typical observation. The fitting
statistics begin to degrade near R=17.
\label{chifit}}
\end{figure}

\begin{figure}[t]
\epsscale{1.00}
\plotone{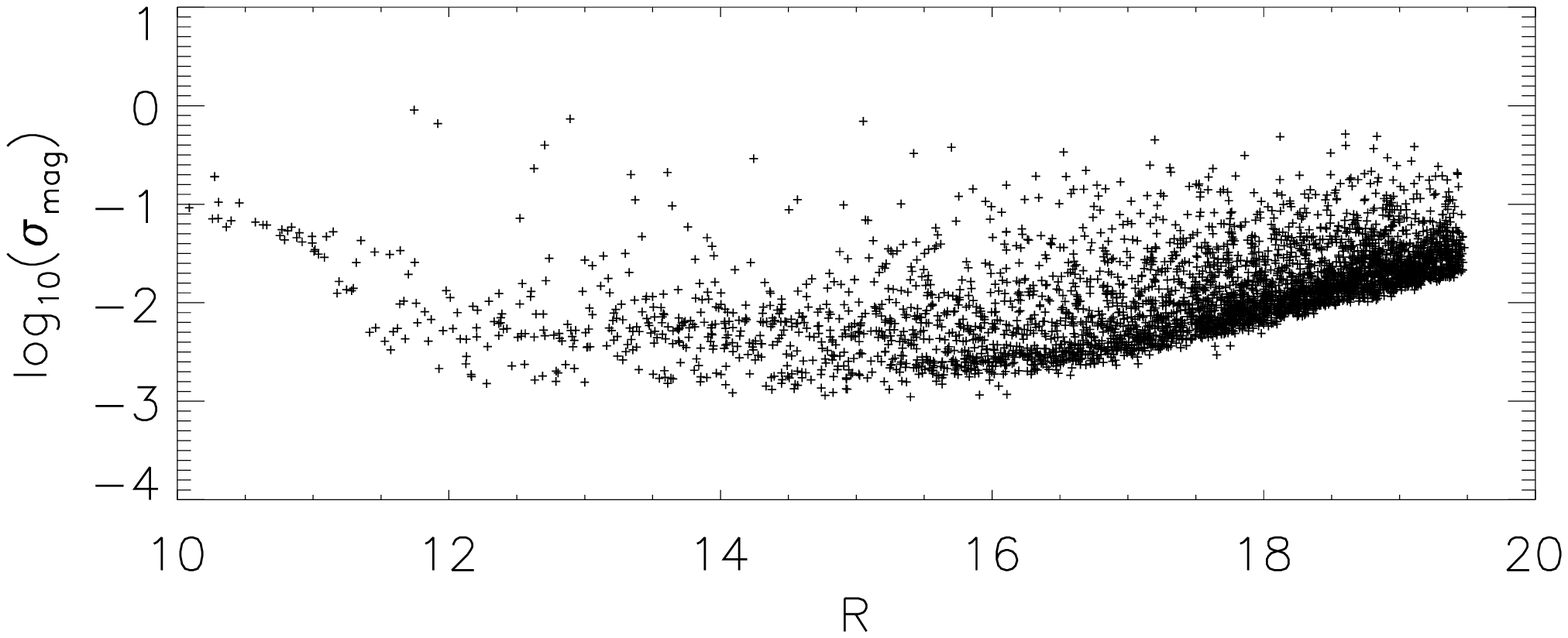}
\caption{Photometric uncertainty for the stars in our sample.
$\sigma$ is the standard deviation of the entire light curve for each point
source. Note the relatively flat error response from R=11.7 to R=16.5. 
\label{magerr}}
\end{figure}

Figure~\ref{chifit} shows $\chi^2$ per degree of freedom
for stars from a typical
observation, illustrating how it worsens for $R<17$, and
Figure~\ref{magerr} shows that the typical uncertainty in the differential
light curves for all our stars bottoms out around
1--2~mmag at $R<17$ rather than continuing to plummet according to
photon statistics. This property of the residual uncertainty is an
important advantage of our square star photometry method. Our best
photometric statistics apply over a flat linear regime covering 4-5 mags
whereas typical CCD photometry follows closely the ``S/N" equation
yielding the best photometry for the top 1-2 mags only and then falling
away (see e.g., \S4.4 in Howell 2000 and Fig. 3 in Everett \& Howell 2001).
The smaller dynamic range in normal CCD observations even holds true for
relatively large defocused images (e.g., Gilliland et al., 1993).

The final step necessary is to bring all of the observations onto a
common zero point, since atmospheric extinction and occasional haze or
cloud obviously change the flux zero point.  This is accomplished by a
similar fit procedure to that of the photometry observations.  All the
observations ($\sim 170$) for a given field are read and all the stars
($\sim 1200$) in that field for which there are flux measurements are
matched up.  For each pair of observations, stars with $12.9<R<16.4$,
which have $\chi^2/N<20$, and which have few masked pixels are
compared with the median magnitude difference and quartiles
contributing to antisymmetric and symmetric matrices.  These are again
analyzed as pairwise differences of an offset vector, and the offset
vector is applied to bring the observations into agreement.  This is
done once for each entire field; stars which have quartile variability
greater than 2.5~mmag are deleted; and the fit is redone for each
quadrant (amplifier) of the OPTIC field.  The final zero point is set
by the photometric observation list.
% Steve: No it's not, at least not for this paper.  Wait until Paper II.
% Our list of instrumental magnitudes is fed into the software used to
% ensemble differential photometry as discussed in Everett \& Howell (2001).

\begin{figure}[t]
\epsscale{1.00}
\plotone{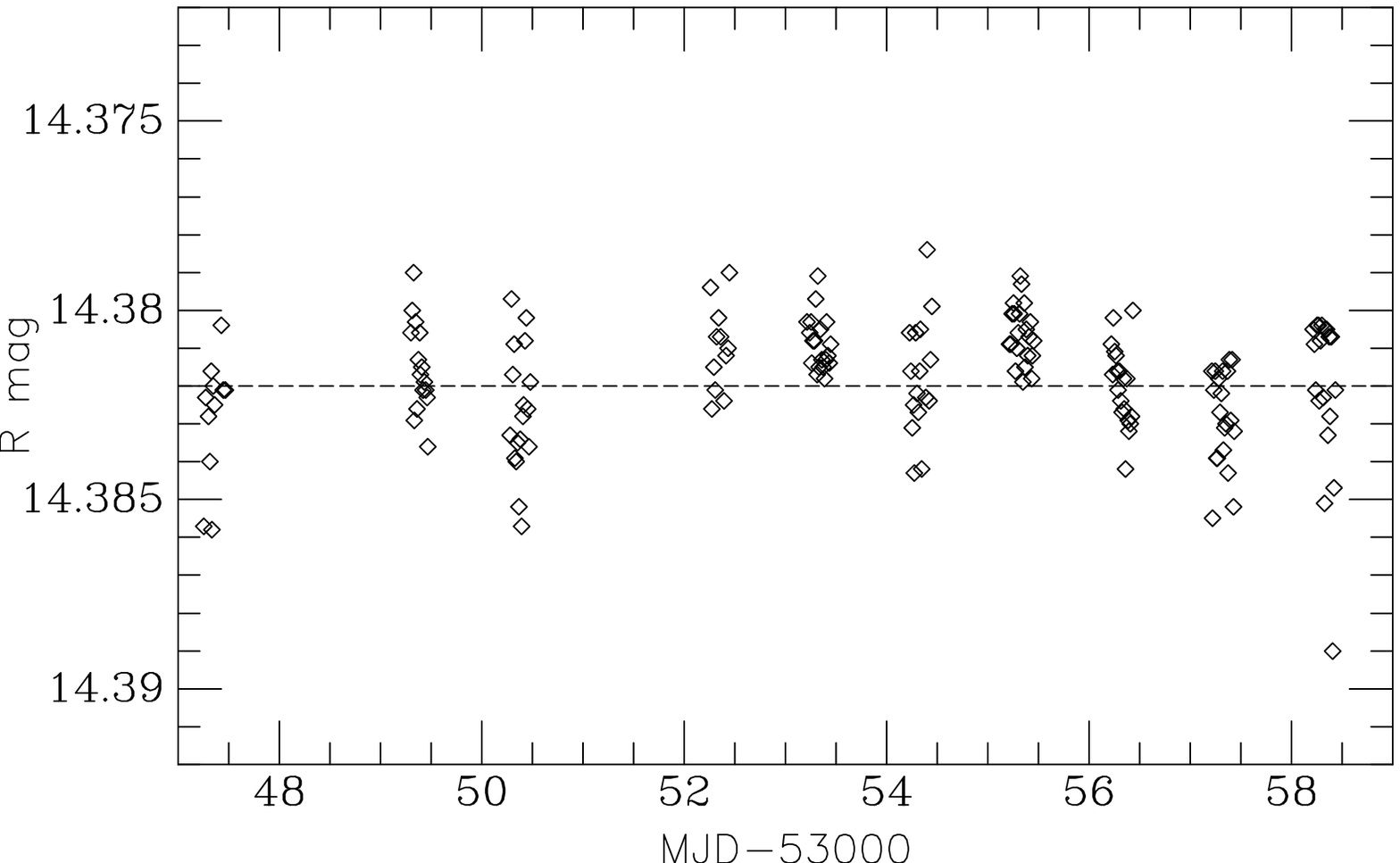}
\caption{Light curve for an $R=14.4$ star for which we do not see
statistically significant variability.  The uncertainties for this star
are 1$\sigma$$\sim$0.0032 mag,
comparable to the scatter, with apparent low-level
systematics visible from day to day.
\label{starc0110}}
\end{figure}

\begin{figure}[t]
\epsscale{1.00}
\plotone{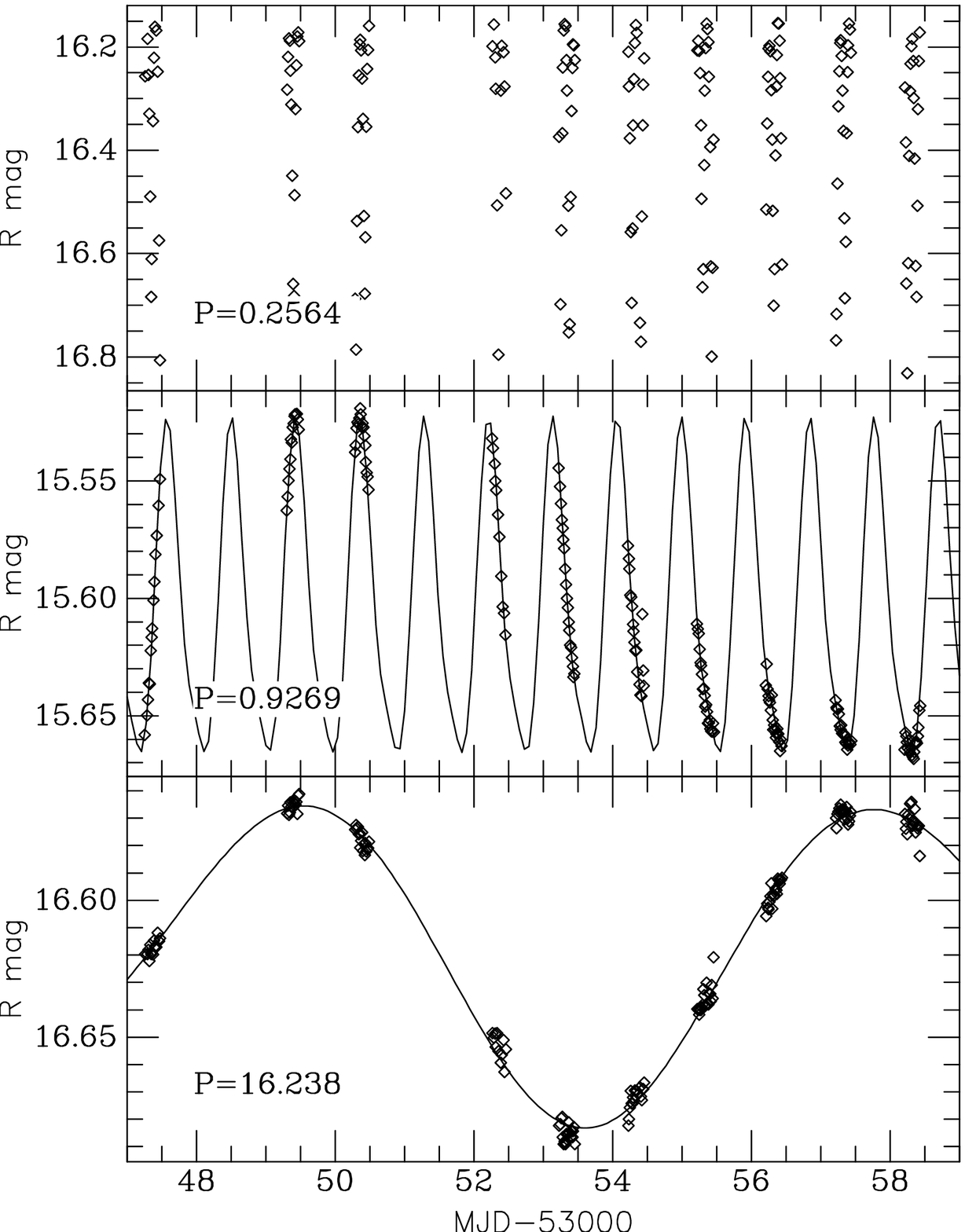}
\caption{Light curves for three variable stars spanning a range in
period (given in units of days).
\label{starvary}}
\end{figure}

Figures~\ref{starc0110} and \ref{starvary} illustrate what the final data
look like.  Figure \ref{starc0110} shows a star in the middle of our
magnitude range which has little or no statistically significant 
variability which we can detect.
We believe this is representative of what our data can reveal when a
star's luminosity is constant.  The half-range between quartiles for
the magnitude distribution (hereafter called ``quartiles'',
approximately a factor of 1.5 smaller than $1\sigma$ for a Gaussian
distribution) is 1.0~mmag.

\begin{figure}[t]
\epsscale{1.00}
\plotone{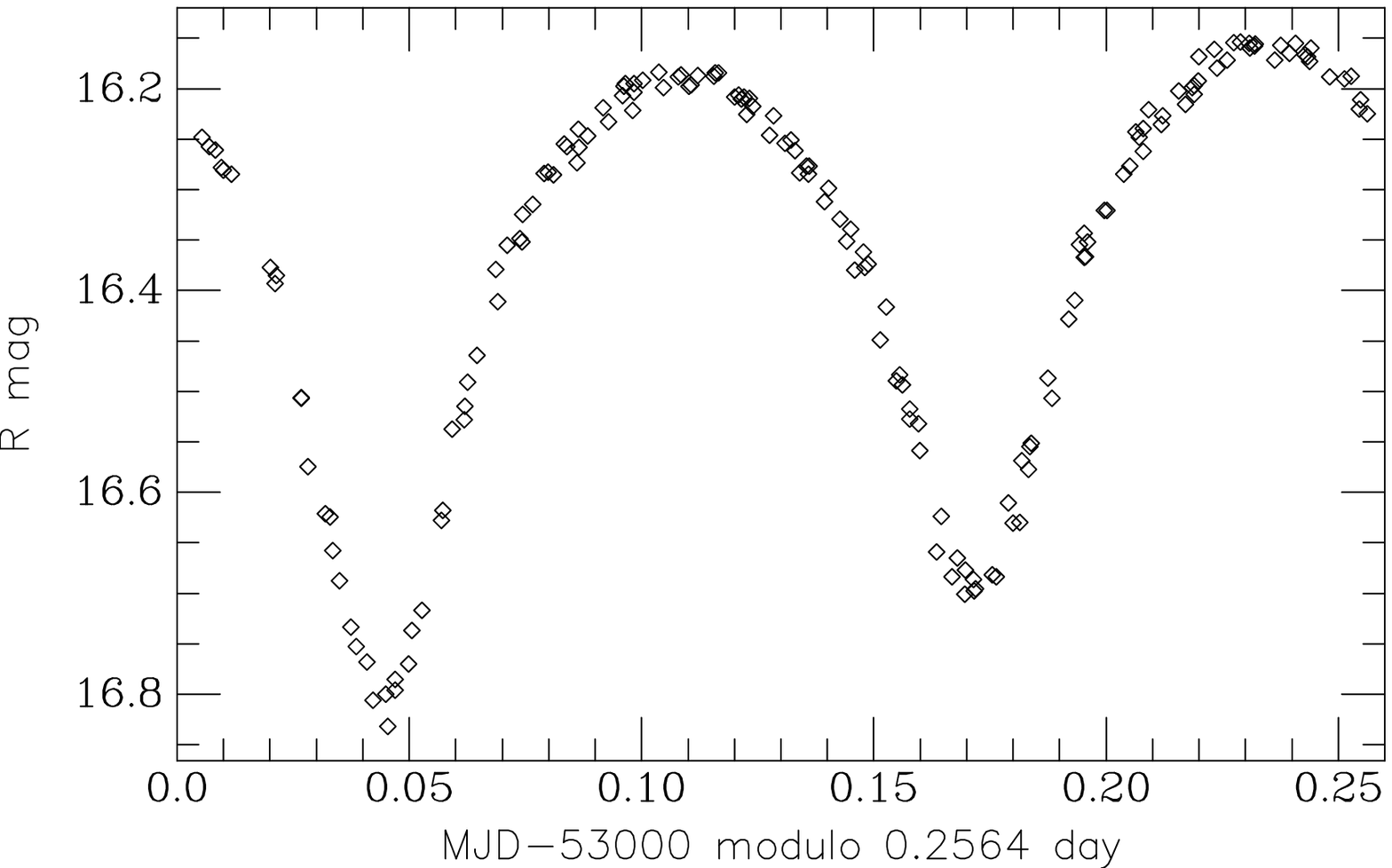}
\caption{Light curve for the short period (contact) eclipsing binary 
star (top of
Figure~\ref{starvary}), phased on its period.
\label{starn0157}}
\end{figure}

Figure~\ref{starvary} shows three variable stars whose periods are
nicely covered by this study.  The quartile residuals of the two
brighter stars are about 2~mmag, and the light curve amplitudes are
evidently approximately 0.06~mag.
Figure~\ref{starn0157} shows the observations of the top star
in Fig.~\ref{starvary} phased around its period.

In order to see if our limiting photometric precision could be due to 
color terms in the data, we performed the following test.
Keep in mind that there are two main steps in our processing 
where uncertainties can creep in. One is the determination of the stellar fluxes and
the other is the use of ensembles of the brightest stars to determine the
point-to-point magnitudes in each light curve.
For our flux determination, we used a single model square star ``PSF"
to determine all the stellar fluxes in each quadrant of the OPTIC 
field of view. We have already discussed
how this was a source of small uncertainties as the detailed shape of the ``top" of the
mesa changed across the field. Given this level of uncertainly, we want to investigate
if the intrinsic color of any given star causes an additional uncertainty in its
brightness determination.

Everett and Howell (2001) discuss this point somewhat but do not present detailed
information to substantiate their findings. The reason is that when making stellar
ensembles to test blue vs. red ensembles used to perform differential photometry on say
red and blue stars, one finds that there are often not enough blue stars in a local
region and using a larger part of the field of view introduces additional uncertainty.
In addition, the fainter stars are usually the reddest and building a fairly red
ensemble to use on say the reddest stars also has lower precision automatically
as it starts with stars of higher uncertainty. The test we tried here is to 
use the likely random placement of stars of various colors to build local ensembles
across the entire field of view of each of the six regions.
We broke up each field of view into 36 boxes in a
6x6 pattern. In each box, an ensemble was chosen consisting of the brightest stars.
Each ensemble was subjected to the tests for ensemble stability as described in Everett
\& Howell (2001). In some cases, the boxes had too few stars
or too few that made good ensemble members so we were forced to 
combine adjacent boxes together. For the more crowded center region (region C)
we ended up with 33 different groups of stars and 15-20 groups in the other
regions. 

Comparison of the resulting light curve uncertainties using this technique with the
light curve uncertainties based on the procedure described above shows 
that the smaller areal ensembles produce, on average, light curves with 10-15\%
better uncertainties. We find, however, no differences in the mean uncertainties
of the light curve variances across the field of view or from group to group.
One assumes that some of the ensembles are likely biased toward the red and are 
used on blue stars and vice verse. Thus, to the best we can tell the use of more local
ensembles, while having the issue of total numbers of good stars available 
and an increased
data reduction effort, allows for slightly better average precisions, but we find no
evidence that differential color terms come into play. 

An additional point in support
of not seeing color term effects is that we do not find the highest level of 
variability to occur in the stars with the most extreme colors (see Paper II). 
Everett et al., (2002) present histograms of variability as a function of 
stellar color and indeed they find a higher percentage of variables at the bluest
($B-V\sim0.25$) and reddest ($B-V\sim1.3$) colors. However, the increase in percent is
marginal at best at the red end (actually more like the mean of the distribution and
only an apparent increase over a low level at $B-V\sim0.5$) and the blue increase 
(which shows larger error bars due to low number statistics) is
due entirely to pulsational variables ($\delta$ Sct and SX Phe types) and not color
terms. 
%Also, blue stars tend to be the brightest in any sample making it is more likely to find smaller levels of
%statistically valid variability. Red stars, on the other hand, tend to be fainter but much 
%more numerous and, in general, more variable, leading to a larger sample of which even equal levels of
%variability will produce numerically more candidates.. 
While the effect of differential color due to atmospheric extinction changes and
over time (airmass) are likely to be present at some level, it appears that to at least
1 mmag they are averaged out through the use of local ensemble differential
photometry. It may be that they are averaged out at even higher precision 
levels as well. Howell \& Tonry (2003) have obtained time series observations 
reaching a precision of 0.5 mmag with no apparent color term variations, although the 
total range of stellar color in the study was smaller than that used here.

\section{Analysis}

For the purpose of finding planets and identifying variability, we
assembled some statistics for each light curve.  The first is simply
the median magnitude and quartiles.  The second pair of statistics are
the ``low-pass'' and ``high-pass'' variability indices.  The first is
the quartiles of the median magnitude for each day, which will flag
light curves which vary on timescales of days or longer.  The second
is the median of the quartiles for each day considered separately,
which will flag light curves which vary significantly within a day.
The third pair of statistics are designed to identify planets.
``extreme'' is the most extreme excursion of a day's median from the
overall median.  ``badday'' is the quartile for the single day with
the largest quartile variability.  The planet search strategy is to
find a star with small ``high-pass'' but with a big ``badday'' which
might signify a 15~mmag eclipse which took place.

\subsection{Occulations}

Experiments with inserting a 15~mmag eclipse 
demonstrated that such an event would clearly stand out in our
statistical tests.  We were particularly interested in periods of
$\sim4$ days since they would potentially give us the opportunity of
seeing more than eclipses, and a Jupiter-sized planet would give us a
15~mmag dip, so we inserted eclipses corresponding to orbits of radius
$\sim0.06$~AU with random phasing into light curves.
A shortcoming to our ``1/2 night" survey is that by covering
only quarter days during this two week
observing run, something like 1/2 to 2/3 of planets with $\sim4$~day
periods would simply fall between the cracks.  Nevertheless, with
favorable phasing a planet causing a 15~mmag eclipse every few days
would stand out very clearly in the parameter-parameter distributions
described in the previous section.  The rank of such artificial
eclipses in terms of separation from the mean parameter correlations
was around the 99th percentile, so it was a simple job to examine such
light curves by eye and decide whether the event looked like an
eclipse or not.  All artificially inserted eclipses were readily
identified. 

However, examination of all light curves with potentially real
planetary occultations revealed none which was a good transit
candidate, at least at the 10~mmag level.  A zero discovery rate is
not out of line with present-day extra-solar planet statistics and
estimates of their discovery rate (Howell et al., 1999) given our
short observing time interval.  An interesting reason for rejection of
most is the fact that the variations in magnitude occurred primarily
on time scales much shorter than those associated with the expected
transits.  The sample we have in this study contains mostly solar type
stars and of these, many appear to be fairly active (rotational and
spot modulations), thus possibly presenting a challenge to transit
hunters.

Carpano et al. (2003) and Jenkins (2002), and others, 
discuss methods of searching for planetary transit signals amoung
highly variable light curves, specifically those associated with active or
rapidly rotating solar-like stars. 
These authors bascially convolve matched filters
(i.e., known transit shapes, depths, and durations) with pre-whitened (often
high S/N) light
curves. The methods they use, detailed filtering
of the light curves to remove frequencies which contaminate them but are
unrelated to transits, plus the possible additional use of color 
information appear to be possible, in theory, for transit detection 
(although yet to be realized in actual data). 
However, our dataset, and most obtained at present by planet hunters, 
has neither the 
time sampling, time duration, or color information 
(or a combination of these) to try such techniques.

\subsection{Stellar variability}

\begin{figure}[t]
\epsscale{1.00}
\plotone{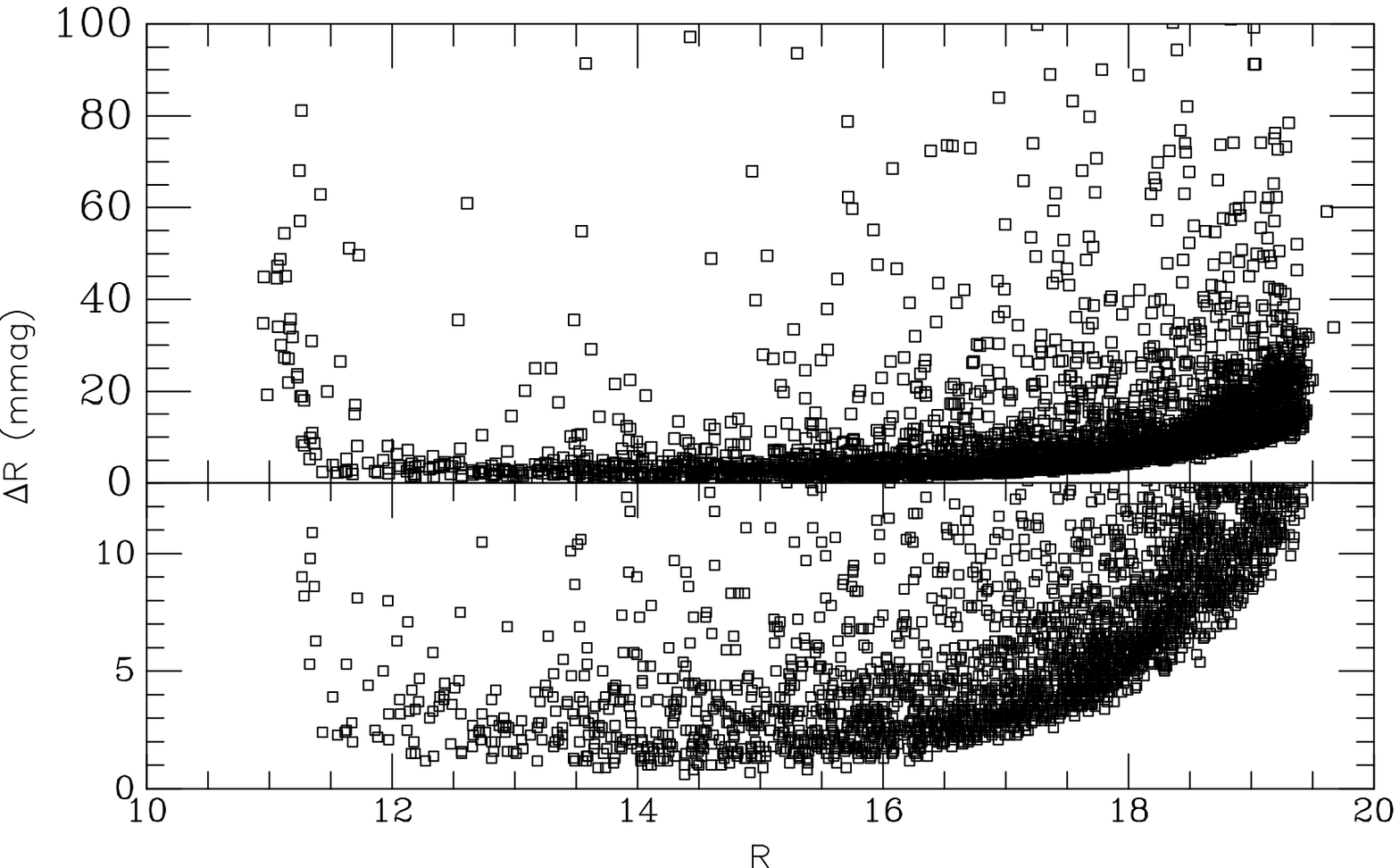}
\caption{Quartile variability for all stars as a function of their $R$
  magnitude.  The top and bottom panels show the same points at
  different scales. We again see the large dynamic range of ``good" data and
the eventual degradation starting near R=17.
\label{quartile}}
\end{figure}

Figure~\ref{quartile} illustrates the variability we see in the light
curves of all the stars for which we measured time series fluxes.  
The abrupt upturn in variability at $R<11.5$ signifies the
onset of saturation, and the errors from photon statistics become
apparent at $R>17$. At the 3$\sigma$ level, we find that 56\% of all our
point sources (over all magnitudes) are statistically variable. While a small percentage of these
variables may be due to remaining low-level systematic effects, we believe that
a 3$\sigma$ cutoff is a conservative estimator.
Of all the variables, 64\% (at a 99\% confidence level)
are periodic with periods from $\sim$20 minutes to many days. This number was
determined by applying a period fitting routine (PDM; Stellingwerf 1978) to 
each total light curve and listing a star as variable if it had a determined
period at the 99\% or higher confidence level. The full details
of the variability nature of NGC 2301, how it manifests itself in
terms of stellar type and brightness and location within the CMD, 
will be discussed in paper II.

For the purposes of more fully understanding the statistics of stellar
variability in this cluster and in our survey, we work with a smaller
sample, restricted to the 5 magnitudes between $12<R<17$.  Both
Fig~\ref{magerr} and Fig~\ref{quartile} assure us that our
observational errors in these regions are approximately 2~mmag or
better.  There are 1148 out of 4462 stars in our sample which meet
this criterion.  (Restricting the data to the 661 stars with $12<R<16$
does not affect the conclusions below.)

%John, I think most of this is best left for paper II
%{\bf lots to say about lots of examples.  Not sure what we want.  I
% find that the ``low-pass'' ``high-pass'' is interesting.  Just about
%  everything with ``low-pass'' $>$ ``high-pass'' is typically a textbook
%  variable star LC, but ``low-pass'' $<$ ``high-pass'' are mostly
%  grumbly stars, and few have short period light curves like C0886.
%  C0191 is a pretty example of P=0.8 hour A=20mmag.}

We note here that typically stars with ``low-pass'' $>$ ``high-pass''
variability are textbook
variable stars with periods of days to weeks.  Some stars with
``low-pass'' $<$ ``high-pass'' have periodic light curves with periods
which are a fraction of a day, but for the most part they show no
obvious periodicity, sometimes with episodic flares or eclipses, but
usually just ``grumbly'' light curves which wander around with
quartiles at the 5-10~mmag level.

\begin{figure}[t]
\epsscale{1.00}
\plotone{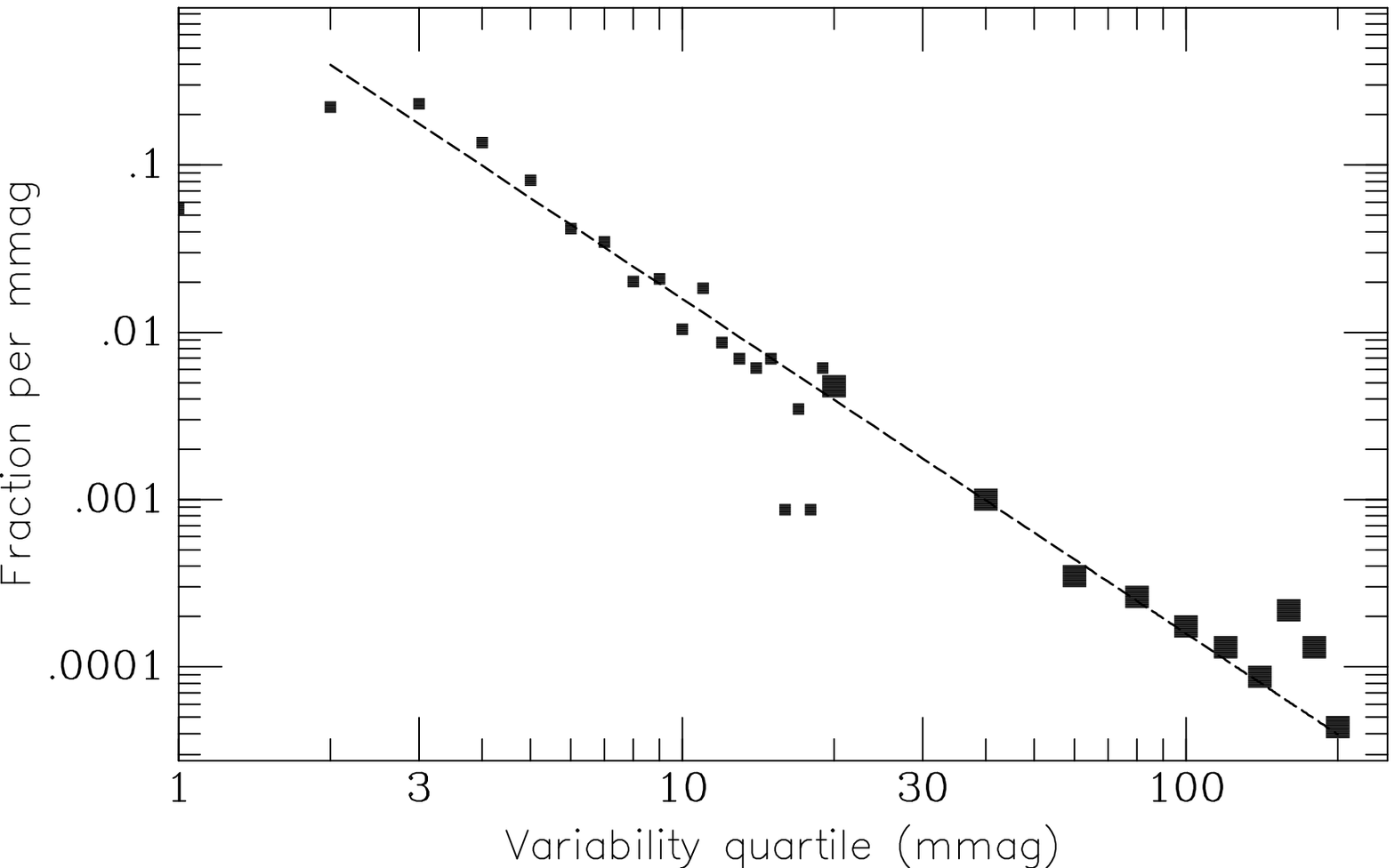}
\caption{Fraction of the sample displaying variability is shown as a function
  of quartile variability amplitude.
  The small points are variability bins of 1~mmag; the
  large points are variability bins of 20~mmag; the straight line is a
  power law of slope $-2$.
\label{vari}}
\end{figure}

Figure~\ref{vari} illustrates the overall level of quartile
variability among this sample of stars.  The data are well described
by a power law of slope $dN/dx = x^{-2}$, where $x$ is the quartile
variability.  This $N(x)$ diverges at $x=0$, of course, but if
restricted to $x>1.6$~mmag the cumulative fraction of stars with
quartile variability $x$~mmag or less is
\begin{equation}
f(<x) = 1 - {1.6~\hbox{mmag}\over x}
\end{equation}
% \begin{equation}
% P(x) = 2 \left(x\over\hbox{1~mmag}\right)^{-2};\qquad\qquad(x>2~\hbox{mmag})
% \end{equation}

The distribution shown in Figure~\ref{vari} reveals a surprising level
of variability among stars. 
(Recall that this is quartile
variability, 1.5 times smaller than $\sigma$ for a Gaussian.)  For
example, the median quartile variability amplitude in our data
(including both periodic and non-periodic sources) is 3.2~mmag, and
1/6 of stars have variability greater than 10~mmag.
To place this into perspective, we examine the
percentage of variability found in two other similar time sampled surveys. 
Huber et al. (2004) obtained V band time series data consisting of
1-2 nights of 10 min exposures, 1 week of occasional sampling, and a 1 year
revisit for 23 sq. degrees of sky located at various Galactic
positions. 
Their photometric precision was 
near 0.01 mag at the bright end and 
they found that that 7-8\% of the 
stars they studied (over all magnitudes in the survey) 
were variable. Everett et al., (2002) obtained
3 minute time sampling of a single 1 sq. degree region for 5 nights in a row
with photometric precision near 5 mmag at the bright end. 
They found 17\% of the sources observed to be variable. 
Our survey reached better photometric precision and kept this precision over a
much larger range in magnitude than these previous examples.
Thus it appears that 
the harder you look, the more variability you find.

The distribution of the variability in terms of amplitude, shape of
the variations, and type of star is of interest as it may be a
critical item which limits photometric searches for small 
planet-to-host-star ratios as well as transit detection at all. Exploration of
variability at levels smaller then 1-2 mmag are being pursued by us
and soon projects such as Pan-STARRS and the Kepler Discovery mission
(launch in 2007) will be producing photometric precisions near
0.5~mmag to 0.1~mmag per measurement.  While ``hot Jupiters"
orbiting Sun-like stars are presumed to be easy targets since they
produce 10~mmag eclipses, smaller planets may be harder to discover,
given the observed level of stellar variability.

\section{Conclusion}
The major goal of this observational work was to test a number of new
technological regimes. We ran the entire time-series operation with
essentially no observer intervention using scripted observations,
predetermined field locations and guide stars, and observational starts
from a remote location (Manoa). We also tested the ability of PSF shaping
for high precision time-series photometry within a fairly crowded stellar
environment and found it
to work well. Our photometric results provide a much larger dynamic 
range of
high precision data than ordinary focused, round star observations 
(nearly 5 mags compared to normally 1-2 mags) and
reached toward 1~mmag precisions with the likelihood of better results using
modified parameters. 

This type of detailed variability study will be carried out at a vastly
larger scale by the Pan-STARRS program; thus our work here can be
considered a precursor of sorts.  Baseline parameters for variability
(and constancy) as functions of spectral type, color, etc. as well as
data collection and reduction experience will be essential inputs for
Pan-STARRS projects.

We have confirmed the use of local ensembles for the highest precision
results as put forth by Everett \& Howell (2001) and showed that
image shape over about a 2 arcmin radius is essentially identical
but slight changes are easily noted at 10 arcmin. These changes are 
caused by
the decorrelation of atmospheric distortion with angular separation,
and can be improved by different OTCCD shifting strategies.

Our temporal survey of NGC 2301 has revealed a wealth of information
on the variability of the cluster stars in terms of spectral type,
location in the CMD, type and amplitude of variability, and other
properties.  Exploration of such datasets for other clusters with
varying age, metallicity, and Galactic location are warranted.  As
higher precision data is obtained, the number of sources that vary
goes up.  This work demonstrates that stellar variability exists down
to mmag levels, and the fraction of stars which are variable increases
inversely as the variability level.  Gross planetary eclipses of hot
Jupiters at the 10~mmag level will not be difficult to spot, but the
grumbly nature of low level variability will make Neptune or
Earth-sized eclipses very hard to identify in actual data.

\acknowledgments
We wish to gratefully acknowledge our referee, Michael Richmond, for a
detailed, insightful, and helpful review of this paper.
We thank the UH TAC for the generous allocation of telescope time.
CV's research was supported by the NOAO/KPNO Research
Experiences for Undergraduates (REU) Program which is funded by the 
National Science Foundation through Scientific Program Order No. 3 
(AST-0243875) of the Cooperative Agreement No. AST-0132798 between 
the Association of Universities for Research in Astronomy (AURA) 
and the NSF.

\end{document}